\documentclass[%
 aip,
 amsmath,amssymb,
 reprint,%
]{revtex4-1}

\usepackage{graphicx}
\usepackage{dcolumn}
\usepackage{bm}
\usepackage{comment}
\usepackage[utf8]{inputenc}
\usepackage[T1]{fontenc}
\usepackage{mathptmx}
\usepackage{chemformula}

\begin{document}

\preprint{AIP/123-QED}

\title[Fast and high-yield fabrication of axially symmetric ion-trap needle electrodes via two step electrochemical etching]{Fast and high-yield fabrication of axially symmetric ion-trap needle electrodes via two step electrochemical etching}

\author{Nikhil Kotibhaskar}
\email{nkotibha@uwaterloo.ca}
\affiliation{ 
Institute for Quantum Computing and Department of Physics and Astronomy, University of Waterloo, Waterloo, ON, Canada
}%
\author{Noah Greenberg}%
\author{Sainath Motlakunta}
\author{Chung-You Shih}
\author{Rajibul Islam}

\date{\today}

\begin{abstract}
Despite the progress in building sophisticated microfabricated ion traps, Paul traps employing needle electrodes retain their significance due to the simplicity of fabrication while producing high-quality systems suitable for quantum information processing, atomic clocks etc.
For low noise operations such as minimizing `excess micromotion', needles should be geometrically straight and aligned precisely with respect to each other.
Self-terminated electrochemical etching, previously employed for fabricating ion trap needle electrodes employs a sensitive and time-consuming technique resulting in a low success rate of usable electrodes. 
Here we demonstrate an etching technique for quick fabrication of straight and symmetric needles with a high success rate and a simple apparatus with reduced sensitivity to alignment imperfections.
The novelty of our technique comes from using a two-step approach employing turbulent etching for fast shaping and slow etching/polishing for subsequent surface finish and tip cleaning.
Using this technique, needle electrodes for an ion-trap can be fabricated within a day, significantly reducing the setup time for a new apparatus.
The needles fabricated via this technique have been used in our ion-trap to achieve trapping lifetimes of several months.

\end{abstract}

\maketitle

\section{\label{sec:level1 introduction}Introduction \protect}

Trapped ions are a versatile platform for Quantum Information processing.
One of the simplest geometries to trap ions, a four-rod Paul trap, employs sharp needles to create confinement\cite{PaulTrapOriginal,FourRodQIP1,FourRodQIP2,Shih2021}.
The surface quality and symmetry of the electrodes determine the usability of such a platform by affecting the ion heating rate\cite{Anomalous1,Anomalous2,Anomalous3,Anomalous4} and the number of ions that can be trapped.
Electrochemical etching is a commonly used technique for the fabrication of nano-sharp tips\cite{STMtips1, STMtips2} used in scanning tunneling microscopes. 
Recently self terminated electrochemical etching has been shown to produce smooth ion trap needle electrodes\cite{Wang2016}.
The complexity of the apparatus and sensitivity of the process to the precise alignment of the components leads to a very low yield. 
Of all the needles we fabricated from this method, only about 20\% were not bent.
Here we present a two-step electropolishing approach that is robust against alignment imperfections and produces smooth, non-bent needles with close to 100\% yield. 
It is this deterministic nature of the technique that allows the fabrication of the ion trap electrodes within a working day.

Figure 1 a) shows a cartoon picture of the trap fabricated in this work.
In a four-rod Paul trap, radio-frequency voltages on the rods create an oscillating saddle potential leading to an effective confining pseudopotential in the x-y direction.
The needles provide the confinement along the z direction through the application of DC voltages.
In state-of-the-art quantum information processing (QIP) experiments, the ions need to be cooled to the lowest quanta of vibration and the axial symmetry of the trap and surface roughness of the electrodes play a crucial role in lowering the heating rates in a trap.
As explained in figure 1 b), the axial symmetry minimizes the excess micromotion by matching the minimum of the RF and DC potentials.
The surface roughness of the electrodes seems to play an important role in the minimization of the so-called `anomalous-heating'\cite{AnomalousRed1,AnomalousRed2,AnomalousRed3}.
The needle electrodes produced with our technique are both axially-symmetric and have high surface quality.
Further by splitting the technique into 2 steps: fast shaping and subsequent polishing, we achieve a reduced fabrication time while retaining the surface quality.
Unlike previous attempts, we exploit the turbulent yet controllable regime of electroethching for speeding up the initial shaping.
The subsequent polishing has been shown to produce <10 nm root-mean-square (RMS) surface roughness ($R_q$) in such electrodes \cite{Wang2016}.
The needles fabricated in this work have a rod end diameter of 0.5 mm and the technique is most suitable for creating needle electrodes with rod diameters between 0.25 mm and 2.5 mm where mechanical machining of the needle shape is difficult.

\begin{figure*}
    \centering
    \includegraphics[width = \linewidth]{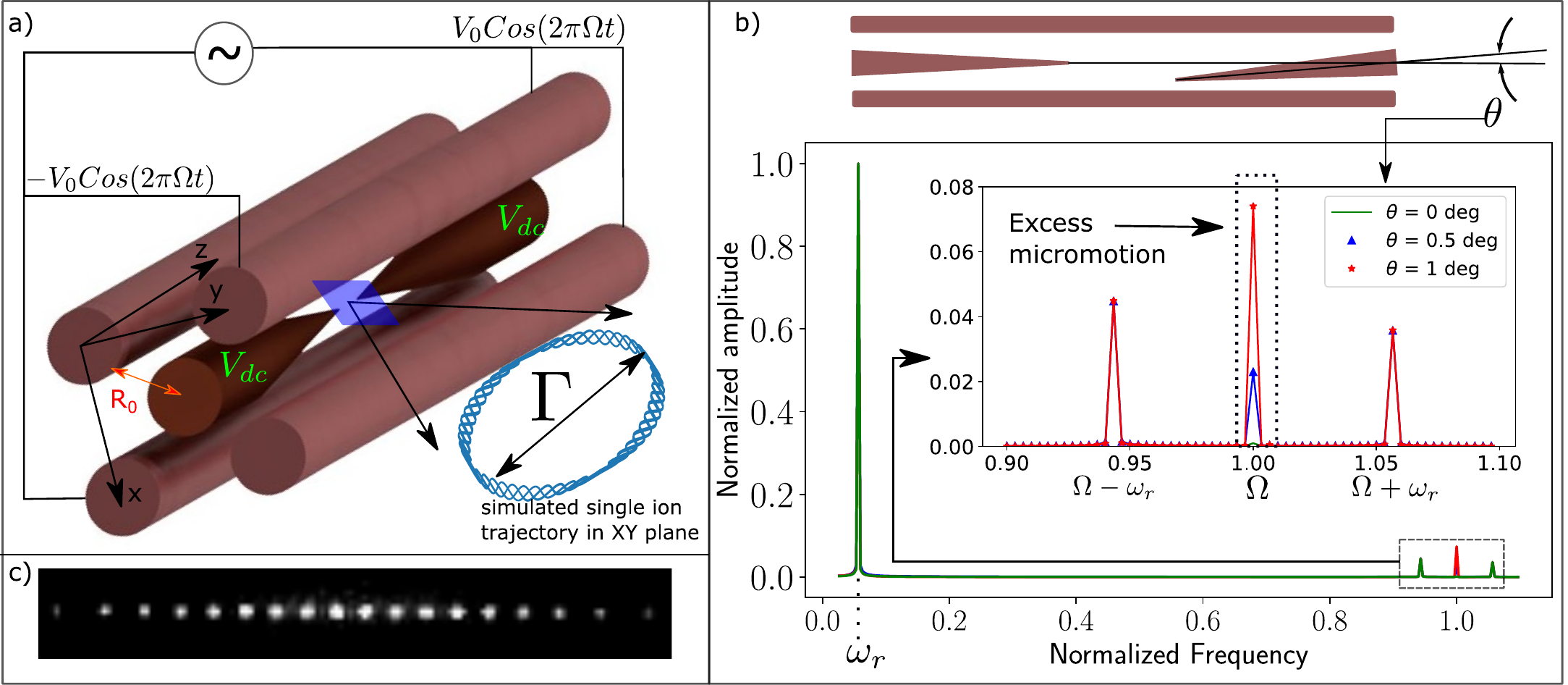}
    \caption{a) Cartoon picture of our trap. Two of the 6 tungsten electrodes are needle electrodes.The simulated trajectory in the inset corresponds to the parameters for our trap i.e drive frequency, $\Omega$  = 20 MHz, amplitude $V_o$ = 250 V $R_o$ = 466 $\mu$m, needle tip-tip distance = 2.8 mm. The extent of the simulated trajectory, $\Gamma$ \~= 40 nm b) FFT of simulated trajectories of a single ion in the trap for different values of angular mismatch between the needles. Notice the peak corresponding to the excess micromotion increases with misaligned/bent needles. We observe that for angular mismatch of greater than 2 degrees, we are unable to obtain stable trajectories even for ions with zero initial velocity placed at the center of the trap. This happens due to a mismatch between the minimum for the RF pseudopotential from the rods and the DC potential from the needles. This mismatch needs to be carefully compensated by DC offsets on each of the rods. Since the compensation only works at a single point in the trapping volume, trapping large linear chains of ions would be very difficult in a conventional 4-rod trap with bent/misaligned needles  c) A chain of 18 bright $^{174}$Yb$^+$ ions in our trap. }
    \label{fig : ion trap}
\end{figure*}

\section{\label{sec:level1 Tungsten Needle Fabrication} Tungsten Needle Fabrication }

In an electroetching process, a tungsten rod (anode) is dipped in an electrolyte (NaOH), and a voltage is applied between the graphite cathode and the anode.
 Tungsten is dissolved into the solution in the form of tungstate ions ($\rm{WO}_4^{2-}$).
 This can be seen from the following reactions\cite{reactions}:

\ch{6 H_2O + 6 e^- -> 3 H_2 + 6 OH^-} \hspace{0.1\linewidth} at the cathode

\ch{ W + 8 OH^- -> WO_4 ^{2-} + 4 H_2 0 + 6 e^-} \hspace{0.1\linewidth} at the anode

\ch{ W + 2 OH^- + 2 H_2 0 -> WO_4^{2-} + 3 H_2} \hspace{0.1\linewidth} overall

\begin{figure*}
    \centering
    \includegraphics[width =1.0
    \linewidth]{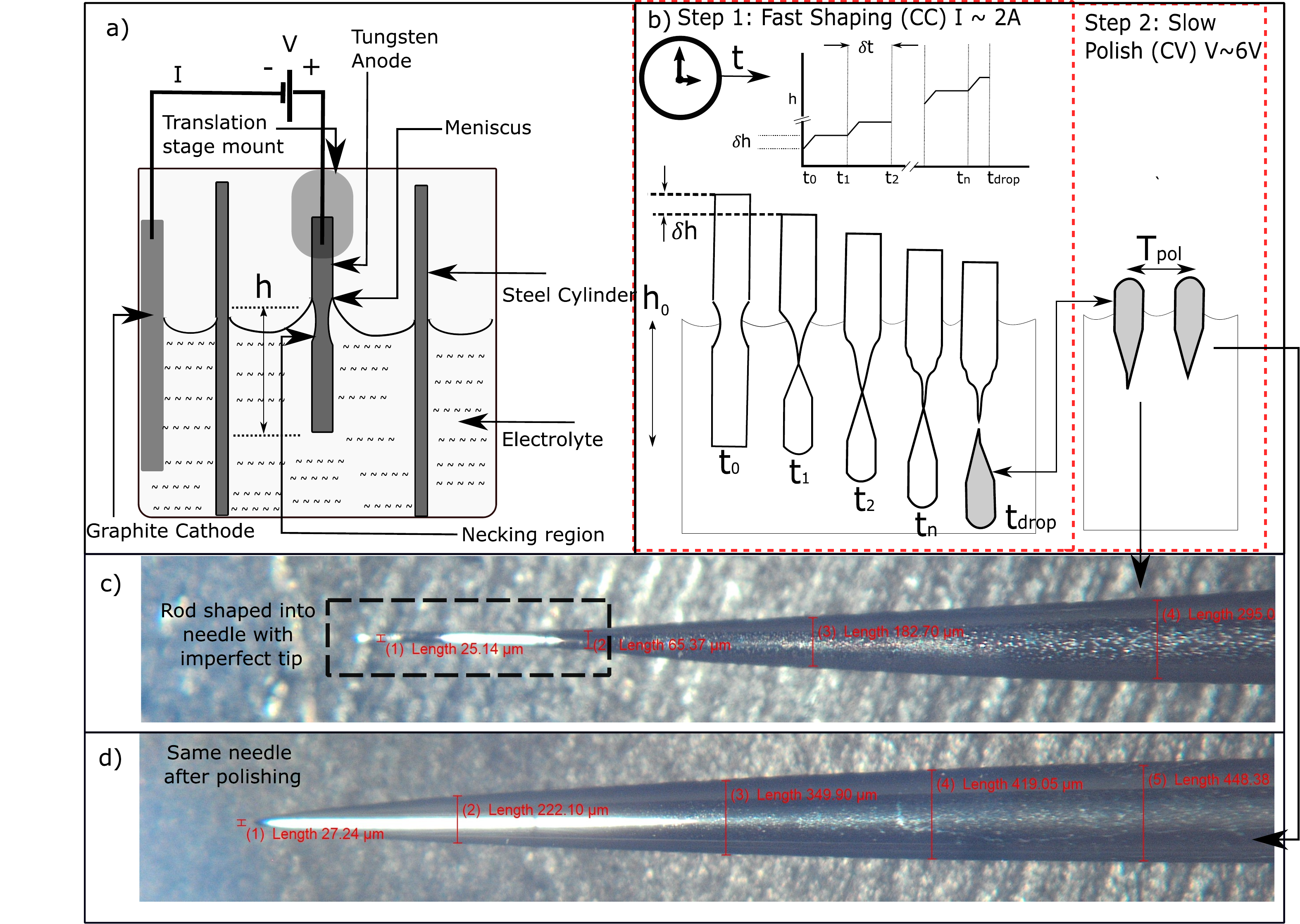}
    \caption{a) Apparatus required for needle fabrication b) Two-step fabrication recipe in this work. 
    The first step is a fast shaping step where the tungsten rod is drawn into the electrolyte intermittently in n steps. 
    The initial current density is set such that the process happens in the turbulent regime. 
    Notice that the neck-like features appear on the top part and the bottom part has a gradual taper with a sharp tip that may or may not be bent. 
    In step 2 the dropped-off part is retrieved from the solution and polished (tip side down) at a constant voltage in the polishing regime c) One of the fabricated needles before step 2 d) the Same needle after step 2}
    \label{needle_polishing_basic_idea}
    
\end{figure*}

The formation of needles from the rods using electroetching is a well-known process\cite{STMtips1}. Fig \ref{needle_polishing_basic_idea} a) shows the formation of a meniscus around the rod near the surface of the electrolyte.
The region below the meniscus, the ``necking region",  has the highest etching rate which can be understood as follows.
Above the necking region, the low concentration of $\rm{OH}^-$ ions reduces the etching reaction rate. 
Below the necking region, the $\rm{WO}_4^{2-}$ ions (percolating down under gravity) stick to the rod and reduce the etching rate. 
At the necking region, a vortex flow\cite{DropOffProcessVortex,STMtips2} preferentially clears away the $\rm{WO}_4^{2-}$ ions from the surface. 
This reduction of reaction rate, above and below, leads to a preferential etching at the 'necking region'.
As etching progresses, the necking region becomes insufficient for sustaining the weight of the rod, and the rod breaks off.
This stops the etching for the bottom piece, which falls into the solution under gravity and has a geometry suitable for use as ion trap electrodes (the top piece is discarded).

The process of electroetching can happen in one of the 4 regimes: Etching, Transient, Polishing, and Breakdown/Turbulent \cite{Wang2016}, depending on the current density and hence the applied voltage.
The polishing regime achieves the best surface quality and has been employed\cite{Wang2016} to fabricate ion trap needle electrodes.
In this technique, a rod is slowly drawn out of the electrolyte while the polishing process continued until drop-off.
We observe that it takes about 45 minutes for a 1 mm diameter tungsten rod to break off in the polishing regime.
Moreover, in this technique, the needle tip profile is strongly influenced by the angle of the needle to gravity.
In practice, the uncertainties in the alignment result in a low yield of usable ion trap electrodes.
In our trials with many tens of fabricated needles our yield was at most 20 \% for non-bent tips.
The slow speed of fabrication in the polishing regime, together with the low yield due to misalignment, results in a slow system development cycle.

In our technique, we divide the fabrication process into two regimes, a fast drop-off in the turbulent regime followed by the conditioning of the dropped-off needle in the polishing regime.
Figure \ref{needle_polishing_basic_idea} b illustrates this process.
Unlike the previous technique, \cite{Wang2016} we drive the needle into the solution instead of drawing it out.
We empirically find that the needle taper is less sensitive to the exact draw rate if it is driven into the solution instead of drawing out, and stepped driving of the needle is sufficient to achieve a gradual taper.
We then etch the dropped-off needle, tip side down, in the polishing regime to clean its tip and achieve a smooth surface finish.
Following this technique, we find that the yield of usable ion trap electrodes is almost 100\% ( see section \ref{sec:level1 Results and Discussion}).

\section{\label{sec:level1 Procedure and Apparatus}Procedure and Apparatus}
The following sections elaborate on the two steps of the fabrication process and how to tune the process parameters. 

\subsection{\label{sec:level2}Needle Shaping (Turbulent regime) }
First, the tungsten rod is mounted on a translation stage and dipped into the electrolyte (2M NaOH solution). 
We use a 3" cylindrical shield \cite{Wang2016} around the anode (tungsten rod) and the graphite cathode is placed outside the shield. 
The cathode and anode are connected with a 30V/3A CC/CV supply which is operated in the constant current (CC) mode to maintain the turbulent regime. 
As etching progresses, we intermittently drive the rod into the solution to obtain a gradual tapering profile on the bottom needle.
While the necking region sees a maximum etching rate the rest of the rod also etches at a lower rate.
Hence, we start with a larger diameter rod than the target non-tip end diameter of the needle electrode.
The following parameters need to be optimized in this step:

\subsubsection{Current Density}
Since the rod is continuously etched, the current density (in the CC mode) increases with time.
In the following, we report the initial current density $J_0$.
Wang, et al (2016) \cite{Wang2016} report that the turbulent/breakdown regime occurs when the current density exceeds 4 mA/mm$^2$.
For a fast drop-off, we operate the etching at the highest $J_0$ for which the process is still controllable.
For a 1 mm diameter rod with immersion depth, $h$ = 45mm  (see Fig \ref{needle_polishing_basic_idea}a) we empirically find a maximum $J_0 \approx$ 14 mA/mm$^2$.
This results in a drop-off time of about 6 minutes compared to 45 minutes at 6 mA/mm$^2$.
Beyond this current density, the solution becomes too effervescent and  $J_0 \approx$ 20 mA/mm$^2$  results in visible shaking of the bottom part.

\subsubsection{Drive Characteristics}
The drive characteristics (See Fig. \ref{needle_polishing_basic_idea} b) determine the sharpness of the needle taper.
As mentioned before, only stepped driving of the rod, by $\delta h$ in each step, (see Fig. \ref{needle_polishing_basic_idea} b) is required to achieve a gradual taper.
The length of the taper, l, is approximately equal to the total extent of the drive into the solution.
For target $l\approx$ 5 mm in our ion trap needles, we find that to achieve a smooth taper the number of steps, $n$, should be between 8 and 15.
After choosing $n$, we estimate $\delta h$ using $\delta h = l/n$ and $\delta t$ using $\delta t = T/n$ where, T is the drop-off time for the specific $J_0$ chosen in the last section.
For our 1 mm diameter rod, we used  $\delta h=0.5$ mm and $\delta t=30 s$.
In each drive, the drive rate is approximately 50 $\mathrm{\mu}$ m/s, i.e the rod is driven 0.5 mm in the first 10 s and left stagnant for the remaining 20 s.
We find that performing the stepped etching with $\delta h$ and $\delta t$ as estimated above leads to the drop-off within a step or two of the chosen number ($n$) of steps.
For $\delta h$ > 0.75 mm multiple meandering features begin to appear on the dropped-off needle instead of a gradual taper and it is recommended to keep $\delta h$ < 0.75 mm for 1-2 mm diameter rods.

\subsection{Step 2 : Polishing }
An example of a dropped-off needle is shown in Fig \ref{needle_polishing_basic_idea} c).
The exact shape of the tip is dependent on the drop-off dynamics which are not precisely controlled.
In addition to the tip defect, the surface roughness of the needle is also high due to the turbulence in the shaping step.
To get rid of the asymmetric part of the tip (shown in the dotted box in Fig 2c) and polish the rest of the needle we etch the needle further in a slow polishing regime.
The dropped-off needle from the shaping step is picked up and placed back into the solution, tip side down, with the full length of the taper immersed into the electrolyte.  
This time the power supply is instead operated in the constant voltage mode. 
The only parameter to be optimized here is the polishing/etching voltage, $V_{\rm{pol}}$. 
We follow Wang et al \cite{Wang2016} to stay in the polishing regime (\textasciitilde6V) in this step.
It is important to note that the surface finish does not improve after $T_{\rm{pol}}$ \textasciitilde2 minutes of polishing. 
The actual polish time may be determined by other constraints. 
For example, we used this step as a means to remove material from the dropped-off needle so that it fits inside the 0.5mm mounting hole of our trap holder.

\section{\label{sec:level1 Results and Discussion}Results and Discussion}
In our ion trap with symmetric needles fabricated as per the recipe above, we can trap a long chain of ions ( see Figure 1 c).
We routinely achieve single ion lifetimes of several days with the longest observed lifetime of 4 months and use the ion for experiments with low error rates suitable for QIP\cite{Shih2021}.
In this section, we discuss the replicability of this technique and the straightness of the fabricated needles.
Tweaking of the recipe for different rod diameters is also discussed. 

\begin{figure*}
    \centering
    \includegraphics[width =1.0\linewidth]{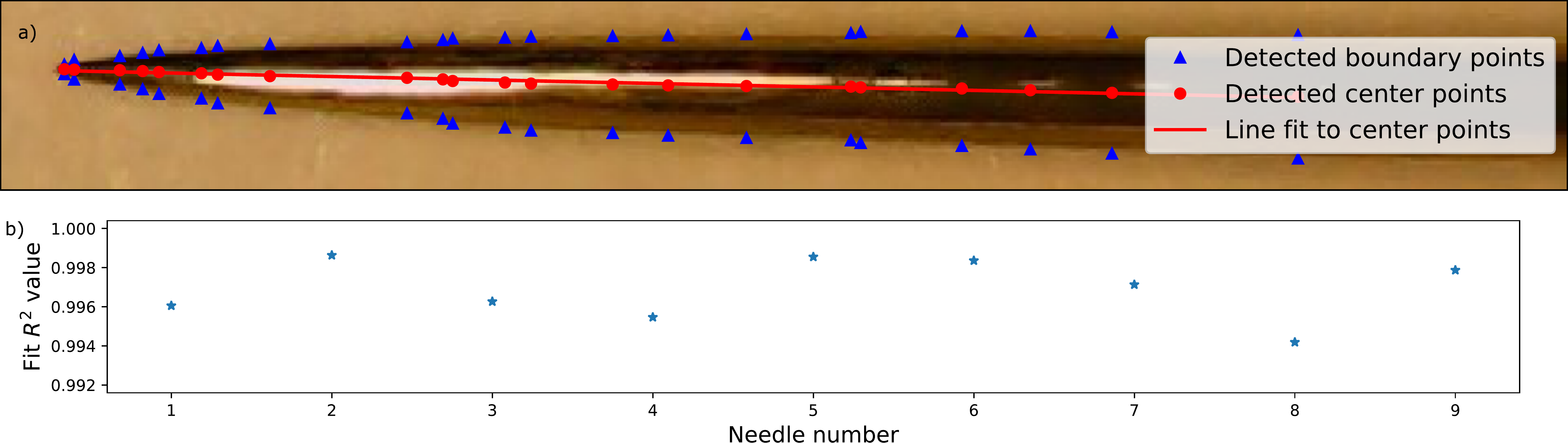}
    \caption{Straightness test for needles a) Centerline-detection: Image processing techniques used to extract boundary and center-line of a needle. b) R$^2$ value from fitting center-line to a straight line for 9 of the 10 needles(one needle damaged while imaging). High R$^2$ values show a good fit and indicate that the needles are indeed straight. Note: to use R$^2$ as a goodness of fit indicator, we ensured that the center-line was not horizontal (leading to a low R$^2$ and falsely suggesting a bad fit) by rotating the images by 10 degrees on each needle.}
    \label{needle_straightness_test}
\end{figure*}


\subsection{Replicability}
To test the replicability, we fabricated 10 needles from commercially available, 1mm-diameter, pure tungsten welding electrodes (McMaster-Carr part no. 8000A521). 
The diameter of the rods was measured to be between 0.95 mm and 1 mm. 
The driving process was manual i.e the micrometer-screw drawing the needle into the solution was turned by hand. 
The use of a motorized translation stage for driving needles was avoided to test the replicability of the process with the least technically demanding setup. 
The process parameters for the test were the following $J_0$ = 14 mA/mm$^2$, $h_0$ = 45mm, $\delta h$ = 0.5mm, $\delta t$ = 30s, $V_{\rm{pol}}$ = 6.4V, $T_{\rm{pol}}$ = 15 minutes.
The fabricated needles were imaged on a microscope (with magnification between 2x and 11.5x), to check for tip defects and axial symmetry. 
All the needles made in the test showed axial symmetry without any bending of the tips and were suitable for our ion trapping apparatus.
 
The exact dimensions of the needles, however, showed variation due to both the initial spread of the rod diameters and the inherent uncertainty in the manual drawing. 
The rod end diameters varied between 0.45-0.55 mm.
This was not a challenge for our application but can be mitigated by choosing initial diameters with higher precision and using a motorized translation stage for finer reproducibility.

\subsection{Straightness}
Apart from tip imperfections the needles also were tested for straightness over the entire extent of the taper.
To test this we first imaged each needle so that the entire extent of the needle could be imaged. 
Then, using standard image processing techniques\cite{CompVision} (frequency domain filtering), the boundary and center-line of the needle were extracted as shown in figure \ref{needle_straightness_test}a). 
The extracted center-line is fit to a straight line, with $R^2>0.994$ (Fig. ~\ref{needle_straightness_test}b).

\subsection{ Results with different rod diameters }
It may be required to fabricate needles with rod diameters other than the ones fabricated in this work.
Here, we discuss the range of needle dimensions where this technique would be suitable.
Since polishing parameters do not change with needle size, we will only discuss the needle shaping step.
Reducing the rod diameter, below the 1mm used here, does not pose a challenge.
However, with larger rod diameters the requirement for maintaining  $J_0 \approx$ 14 mA/mm$^2$ leads to higher current requirements in the shaping step causing the solution to be effervescent.
We successfully fabricated needles with rod diameters 0.75 mm, 1.14 mm, and 1.58 mm by optimizing $\delta h$ and $\delta t$ as discussed in section \ref{sec:level1 Procedure and Apparatus}.
We also fabricated needles with 2.4mm diameter rods and found that the solution becomes too effervescent making the process uncontrollable and dropping the yield to about 50 \%.
For larger diameters, $J_0$ can be reduced at the expense of the fabrication time.
It was however noticed that tungsten rods with a diameter greater than 2.5 mm are quite stiff and mechanical grinding would be more suitable to fabricate a conical taper.
The shaped needle can then be polished using the polishing step.
Due to this, we conclude that this technique is most suitable for rod diameters between 0.75 mm and 2.5 mm.

\section{Conclusion}
We have developed a novel technique for fast fabrication of ion-trap needle electrodes with high yield using two-step electrochemical etching. 
The apparatus required for the fabrication is extremely simple and produces smooth, axially-symmetric needles suitable for QIP experiments. 
The technique can be easily tweaked to make needles from rods of diameters different from the ones used here. 
Moreover, the technique affords minimal setup time and needle electrodes for a conventional 4-rod trap can be fabricated within a day. 
This would lead to a significant reduction in the setup time for a new ion-trapping apparatus.

\section{Acknowledgements}
We thank Roland Hablützel and Manas Sajjan for helpful discussions. We thank Joshua Da Costa for testing out the technique by fabricating 10 needle electrodes.
We acknowledge financial support from the Natural Sciences and Engineering Research Council of Canada Discovery (RGPIN-2018-05250) program, Ontario Early Researcher Award, Canada First Research Excellence Fund (CFREF), NFRF Grant, University of Waterloo, and Innovation, Science and Economic Development Canada (ISED).


\nocite{*}
\bibliography{aipsamp}

\end{document}